\documentclass[twocolumn,tighten]{aastex62}
\usepackage{color}

\usepackage{multirow}

\shortauthors{Sano et al.}

\received{May 26, 2018}
\revised{March 10, 2019}
\accepted{March 16, 2019}
\published{April 30, 2019}

\begin{document}

\title{Possible Evidence for Cosmic-Ray Acceleration in the Type Ia SNR RCW~86:\\Spatial Correlation between TeV Gamma rays and Interstellar Atomic Protons}

\author[0000-0003-2062-5692]{H. Sano}
\affiliation{Institute for Advanced Research, Nagoya University, Furo-cho, Chikusa-ku, Nagoya 464-8601, Japan; sano@a.phys.nagoya-u.ac.jp}
\affiliation{Department of Physics, Nagoya University, Furo-cho, Chikusa-ku, Nagoya 464-8601, Japan}

\author[0000-0002-9516-1581]{G. Rowell}
\affiliation{School of Physical Sciences, The University of Adelaide, North Terrace, Adelaide, SA 5005, Australia}

\author{E. M. Reynoso}
\affiliation{Instituto de Astronom${\acute{\i}}$a y F${\acute{\i}}$sica del Espacio (IAFE, CONICET-UBA), Av. Int. G${\ddot{u}}$raldes 2620, Pabell${\acute{o}}$n IAFE, Ciudad Universitaria, Ciudad Aut${\acute{o}}$noma de Buenos Aires, Argentina}

\author{I. Jung-Richardt}
\affiliation{Universit${\ddot{a}}$t Erlangen-N${\ddot{u}}$rnberg, Physikalisches Institut, Erwin-Rommel-Str. 1, D 91058 Erlangen, Germany}

\author[0000-0001-8296-7482]{Y. Yamane}
\affiliation{Department of Physics, Nagoya University, Furo-cho, Chikusa-ku, Nagoya 464-8601, Japan}

\author{T. Nagaya}
\affiliation{Department of Physics, Nagoya University, Furo-cho, Chikusa-ku, Nagoya 464-8601, Japan}

\author[0000-0002-2458-7876]{S. Yoshiike}
\affiliation{Department of Physics, Nagoya University, Furo-cho, Chikusa-ku, Nagoya 464-8601, Japan}

\author{K. Hayashi}
\affiliation{Department of Physics, Nagoya University, Furo-cho, Chikusa-ku, Nagoya 464-8601, Japan}

\author[0000-0003-2450-1611]{K. Torii}
\affiliation{{Nobeyama Radio Observatory, National Astronomical Observatory of Japan (NAOJ), National Institutes of Natural Sciences (NINS), 462-2, Nobeyama, Minamimaki, Minamisaku, Nagano 384-1305, Japan}}

\author[0000-0003-2762-8378]{N. Maxted}
\affiliation{{School of Sciences, University of New South Wales, Australian Defence Force Academy, Canberra, ACT 2600, Australia}}

\author{I. Mitsuishi}
\affiliation{Department of Physics, Nagoya University, Furo-cho, Chikusa-ku, Nagoya 464-8601, Japan}

\author{T. Inoue}
\affiliation{Department of Physics, Nagoya University, Furo-cho, Chikusa-ku, Nagoya 464-8601, Japan}

\author[0000-0003-4366-6518]{S. Inutsuka}
\affiliation{Department of Physics, Nagoya University, Furo-cho, Chikusa-ku, Nagoya 464-8601, Japan}

\author{H. Yamamoto}
\affiliation{Department of Physics, Nagoya University, Furo-cho, Chikusa-ku, Nagoya 464-8601, Japan}

\author[0000-0002-1411-5410]{K. Tachihara}
\affiliation{Department of Physics, Nagoya University, Furo-cho, Chikusa-ku, Nagoya 464-8601, Japan}

\author{Y. Fukui}
\affiliation{Institute for Advanced Research, Nagoya University, Furo-cho, Chikusa-ku, Nagoya 464-8601, Japan; sano@a.phys.nagoya-u.ac.jp}
\affiliation{Department of Physics, Nagoya University, Furo-cho, Chikusa-ku, Nagoya 464-8601, Japan}

\begin{abstract}
We present a detailed morphological study of TeV gamma rays, synchrotron radiation, and interstellar gas in the young Type Ia supernova remnant (SNR) RCW~86. We find that the interstellar atomic gas shows good spatial correlation with the gamma rays, indicating that the TeV gamma rays from RCW~86 are likely to be dominantly of hadronic origin. In contrast, the spatial correlation between the interstellar molecular cloud and the TeV gamma rays is poor in the southeastern shell of the SNR. We argue that this poor correlation can be attributed to the low-energy cosmic rays ($\sim1$ TeV) not penetrating into the dense molecular cloud due to an enhancement of the turbulent magnetic field around the dense cloud of $\sim10$--100 $\mu$G. We also find that the southwestern shell, which is bright in both synchrotron X-ray and radio continuum radiation, shows a significant gamma-ray excess compared with the interstellar proton column density, suggesting that leptonic gamma rays via inverse Compton scattering possibly contributes along with hadronic gamma rays. The total cosmic-ray energies of the young TeV gamma-ray SNRs---RX~J1713.7$-$3946, Vela~Jr, HESS~J1731$-$347, and RCW~86---are roughly similar, which indicates that cosmic rays can be accelerated in both the core-collapse and Type Ia supernovae. The total energy of cosmic rays derived using the gas density, $\sim10^{48}$--$10^{49}$ erg, gives a safe lower limit due mainly to the low filling factor of interstellar gas within the shell.

\end{abstract}

\keywords{cosmic rays --- gamma rays: ISM --- ISM: clouds --- ISM: individual objects (RCW~86) --- ISM: supernova remnants}

\section{Introduction} \label{sec:intro}
A long-standing question is how cosmic rays, which mainly consist of relativistic protons, are accelerated in interstellar space. Supernova remnants (SNRs) are the most reliable candidates for acceleration sites of Galactic cosmic rays with an energy up to $\sim3 \times 10^{15}$ eV (the ``knee'' energy) because high-velocity shockwaves of $\sim3000$--10000 km s$^{-1}$ offer an ideal site for diffusive shock acceleration \citep[e.g.,][]{1978ApJ...221L..29B,1978MNRAS.182..147B}. However, the principal acceleration sites of cosmic rays are still being debated due to lack of sufficient observational evidence.

Young SNRs with bright TeV gamma-ray shells have received much attention on account of their high potential for accelerating cosmic rays close to the knee energy. TeV gamma rays from young SNRs are mainly produced by relativistic cosmic-ray protons and electrons through two mechanisms, called hadronic or leptonic processes. In the hadronic process, interactions between cosmic rays and interstellar protons produce a neutral pion that quickly decays to two gamma-ray photons. Conversely, cosmic-ray electrons energize a low-energy photon to TeV gamma-ray energy via inverse Compton scattering\footnote{Bremsstrahlung of high energy electrons can also explain gamma-ray emission, but in general, inverse Compton scattering is the dominant leptonic process in young SNRs.}. Numerous attempts have been made to distinguish the two processes by using broad-band spectral modeling in the radio, X-ray, and gamma-ray range. In most cases, however, it is difficult to distinguish between hadronic and leptonic gamma rays \citep[e.g.,][]{2018AA...612A...4H,2018AA...612A...6H,2018AA...612A...7H}.

Investigating the interstellar neutral gas associated with SNRs holds a key to distinguishing between hadronic and leptonic gamma rays. \cite{2012ApJ...746...82F} present a good spatial correspondence between TeV gamma rays and the total interstellar proton column density---taking into account both the molecular and atomic components---in the young shell-type SNR RX~J1713.7$-$3946 ($\sim1600$ yr). This provides one of the essential conditions for gamma-rays to be dominantly of hadronic origin, because in such case, the gamma-ray flux is proportional to the target-gas density if we assume an azimuthally isotropic distribution of cosmic rays. By estimating an average interstellar proton density of $\sim130$ cm$^{-3}$, Fukui et al. derived the total energy of accelerated cosmic rays to be $\sim10^{48}$ erg, corresponding to $\sim0.1\%$ of the total kinetic energy released in a supernova explosion. Subsequent studies presented similar results for young TeV gamma-ray SNRs HESS~J1731$-$347 \citep{2014ApJ...788...94F} and Vela~Jr \citep{2017ApJ...850...71F}. All are thought to be core-collapse SNRs; they are expected to be strongly associated with the rich interstellar gas. To better understand the origin of cosmic rays and their energies, we need to look to other gamma-ray SNRs that have a different type of progenitor, such as Type Ia SNRs.

RCW~86 (also known as G315.4$-$2.3 or MSH~14$-$63) is a Type Ia SNR in the southern sky with a bright GeV--TeV gamma-ray shell \citep[e.g.,][]{1989ApJ...337..399C,2007PASJ...59S.171U,2008PASJ...60S.123Y,2009ApJ...692.1500A,2011ApJ...741...96W,2018AA...612A...4H,2016ApJ...819...98A}. The shell diameter is $\sim30$ pc ($\sim40$ arcmin) at a distance of 2.5 kpc \citep[e.g.,][]{2018AA...612A...4H}, which is well suited for morphological studies. The young age of $\sim1800$ yr \citep{1975Obs....95..190C,2006ChJAA...6..635Z} makes it a target for investigating the acceleration of cosmic rays close to the knee energy.

The origin of the gamma ray emission in RCW~86 has been discussed for several years. \cite{2018AA...612A...4H} used spectral modeling to demonstrate that both the leptonic and hadronic scenarios are explained by the observed GeV--TeV gamma rays. Conversely, \cite{2016ApJ...819...98A} used one- and two-zone spectral models but showed only the leptonic scenarios. {This is because the observed GeV gamma-ray emission shows a hard spectra with a spectral index of $\sim$1.42, which is consistent with the leptonic origin under the standard diffusive shock acceleration (DSA) model. The authors also discovered an H{\sc i} void toward the SNR whose velocity is $\sim$35 km s$^{-1}$. However, no detailed study has yet compared the gamma-ray distribution with the interstellar-gas including both the molecular clouds and atomic gas.}

{Subsequently}, \citeauthor{2017JHEAp..15....1S} (\citeyear{2017JHEAp..15....1S}; hereafter Paper I) presented the interstellar molecular and atomic gas distribution toward the SNR RCW~86. They concluded that an interstellar gas with a radial velocity from $-42$ to $-28$ km s$^{-1}$ is likely associated with the SNR based on three elements: (1) the CO and H{\sc i} show a good spatial correspondence with the X-ray shell, (2) H{\sc i} shows an expanding gas motion due to accretion winds from the progenitor system, and (3) an enhanced CO $J$ = 2--1/1--0 intensity ratio only in the surface of the molecular clouds is detected, suggesting heating and/or compression by the shockwaves (see Section 3 in Paper I). They also found an H{\sc i} envelope {on} the molecular cloud, indicating that the progenitor system of RCW~86 had a weaker wind than that of the core-collapse SNR RX~J1713.7$-$3946. This further supports the idea that RCW~86 is a Type Ia SNR, and its progenitor system is a white dwarf and a low-mass star with weak velocity accretion winds.

In this paper, we present a detailed morphological study of the TeV gamma rays, interstellar molecular and atomic gases, and synchrotron radio and X-rays to investigate whether the TeV gamma rays are dominantly from hadronic or leptonic origins. Section \ref{sec:obs} describes the observational datasets. Section \ref{sec:results} comprises four subsections: subsection \ref{subsec:distribution} gives the distributions of the TeV gamma rays; subsections \ref{sec: synchrotron} and \ref{sec: hi} make a spatial comparison of the TeV gamma rays, synchrotron radiation, and CO/H{\sc i}; subsection \ref{sec: ism} presents the total interstellar proton map and compares it with the TeV gamma rays. Discussion and conclusions are given in sections \ref{sec: discussion} and \ref{sec: conclusions}, respectively.

\section{Observational Datasets} \label{sec:obs}
\subsection{TeV Gamma Rays}
We used the TeV gamma-ray image of RCW~86 shown in Figure \ref{fig1} of \cite{2018AA...612A...4H}. The TeV gamma-ray image corresponds to energies above $\sim100$ GeV. The total exposure time was $\sim57$~h for RCW~86. The point spread function (PSF) of the image is 0\fdg061 ($68\%$ radius). The image was smoothed with a Gaussian function with 0\fdg06, resulting in a PSF of $\sim0\fdg086$.

\subsection{{GeV Gamma Rays}}
{The test static (TS) map of GeV gamma-rays appeared in \cite{2016ApJ...819...98A} was also used for tracing the low-energy cosmic rays. The energy range of the TS map is above 1 GeV. The PSF of the image is $\sim$0\fdg27 ($68\%$ radius). A more detailed analysis can be found in \cite{2016ApJ...819...98A}.}

\begin{deluxetable*}{cccccccc}[]
\tablewidth{\linewidth}
\tablecaption{Summary of $XMM$-$Newton$ archive data}
\tablehead{
&&&&&\multicolumn{3}{c}{Exposure}\\
\cline{6-8}\\
Observation ID &  $\alpha_{\mathrm{J2000}}$ & $\delta_{\mathrm{J2000}}$ & Start Date  & End Date & MOS1 & MOS2 & pn \\
& (degree) & (degree) & (yyyy-mm-dd hh:mm:ss) & (yyyy-mm-dd hh:mm:ss) & (ks) & (ks) & (ks)  \\
}
\startdata
0110010701 & 220.73 & $-62.63$ & 2000-08-16 04:04:38 & 2000-08-16 10:43:07 & 16 & 16 & 15\\
0110011301 & 221.31 & $-62.41$ & 2000-08-16 12:03:46 & 2000-08-16 17:37:28 & 11 & 11 & \phantom{0}5\\
0110011401 & 220.51 & $-62.22$ & 2000-08-16 20:18:03 & 2000-08-17 01:36:33 & \phantom{0}9 & 10 & \phantom{0}6\\
0110010501 & 220.14 & $-62.60$ & 2001-08-17 11:47:26 & 2001-08-17 16:25:47 & \phantom{0}9 & \phantom{0}7 & \phantom{0}3\\
0110012501 & 220.24 & $-62.72$ & 2003-03-04 09:46:14 & 2003-03-04 13:11:34 & \phantom{0}9 & \phantom{0}8 & \phantom{0}6\\
0208000101 & 221.26 & $-62.34$ & 2004-01-26 22:30:59 & 2004-01-27 15:12:51 & 46 & 47 & 44\\
0504810101 & 221.57 & $-62.30$ & 2007-07-28 07:45:25 & 2007-07-29 16:12:53 & 94 & 99 & 76\\
0504810601 & 221.57 & $-62.30$ & 2007-07-30 15:45:31 & 2007-07-31 01:52:21 & 19 & 19 & 16\\
0504810201 & 221.40 & $-62.47$ & 2007-08-13 17:42:42 & 2007-08-14 14:37:56 & 50 & 55 & 35\\
0504810401 & 220.15 & $-62.60$ & 2007-08-23 03:17:26 & 2007-08-23 23:33:12 & 62 & 62 & 50\\
0504810301 & 220.50 & $-62.22$ & 2007-08-25 02:49:31 & 2007-08-25 23:34:05 & 61 & 62 & 44\\
0724940101 & 221.22 & $-62.68$ & 2014-01-27 18:48:07 & 2014-01-29 00:03:07 & 95 & 95 & 77\\
\enddata
\tablecomments{All exposure times represent the flare-filtered exposure.}
\label{tab1}
\vspace*{-0.1cm}
\end{deluxetable*}

\subsection{Radio Continuum and X-rays}
We use the 843~MHz radio continuum image obtained by using the Molonglo Observatory Synthesis Telescope (MOST) installed in Australia \citep{1996A&AS..118..329W}. The angular resolution of the radio continuum is $\sim43''$. To create the X-ray image, we used the datasets of $XMM$-$Newton$ presented in Paper I. We used the $XMM$-$Newton$ Science Analysis System (SAS) version 16.0.0 and HEAsoft version 6.18 to analyze both the EPIC-MOS and EPIC-pn datasets with a total of twelve pointings (see Table \ref{tab1}). Because the X-ray emission of RCW~86 fills the field of view of the EPICs, we used the $XMM$-$Newton$ Extended Source Analysis Software \citep[ESAS;][]{2008A&A...478..575K}. We filtered out soft proton flares by using the ESAS procedure, resulting in a good exposure time of 481 ks for EPIC-MOS1, 491 ks for EPIC-MOS2, and 377 ks for EPIC-pn. To obtain quiescent particle background (QPB) images and exposure maps, we run the {\it{mos-/pn-back}} and {\it{mos-/pn-filter}} scripts. The {\it{merge\_comp\_xmm}} script was also used to combine the twelve pointing data. Finally, we applied an adaptive smoothing by using the {\it{adapt\_merge}} script, where the smoothing counts and pixel sizes were set to 150 counts and $6''$, respectively. We obtained QPB-subtracted, exposure-corrected, and adaptively smoothed images in the energy band of 2--5 keV, which is dominated by the continuum radiation from synchrotron X-rays produced by cosmic-ray electrons with TeV energy \citep[][]{2002ApJ...581.1116R,2016ApJ...819...98A}.

\subsection{CO and H{\sc i}}
To estimate total interstellar proton column density in both the molecular and atomic forms, we used the $^{12}$CO($J$ = 1--0) and H{\sc i} data from Paper I. The CO data were acquired by using NANTEN2 installed in the Republic of Chile, while the H{\sc i} data were acquired by using the Australia Telescope Compact Array (ATCA) and combined with single-dish data from the Parkes 64-m radio telescope. The final beam size of CO was $\sim180''$ and that of H{\sc i} was $160'' \times 152''$ with a position angle of $-3\arcdeg$. The typical noise level is $\sim0.42$ K at 0.16 km s$^{-1}$ velocity resolution for CO, and $\sim1.0$ K at 0.82 km s$^{-1}$ velocity resolution for H{\sc i}. 

Paper I analyzes in detail both the CO and H{\sc i} spatial and velocity distributions and conclude that the interstellar gas in the velocity range from $-46$ to $-28$ km s$^{-1}$ is most likely associated with RCW~86. Therefore, in the present study, we adopt the same velocity range as the interstellar gas associated with the SNR.

\begin{figure*}
\begin{center}
\includegraphics[width=\linewidth,clip]{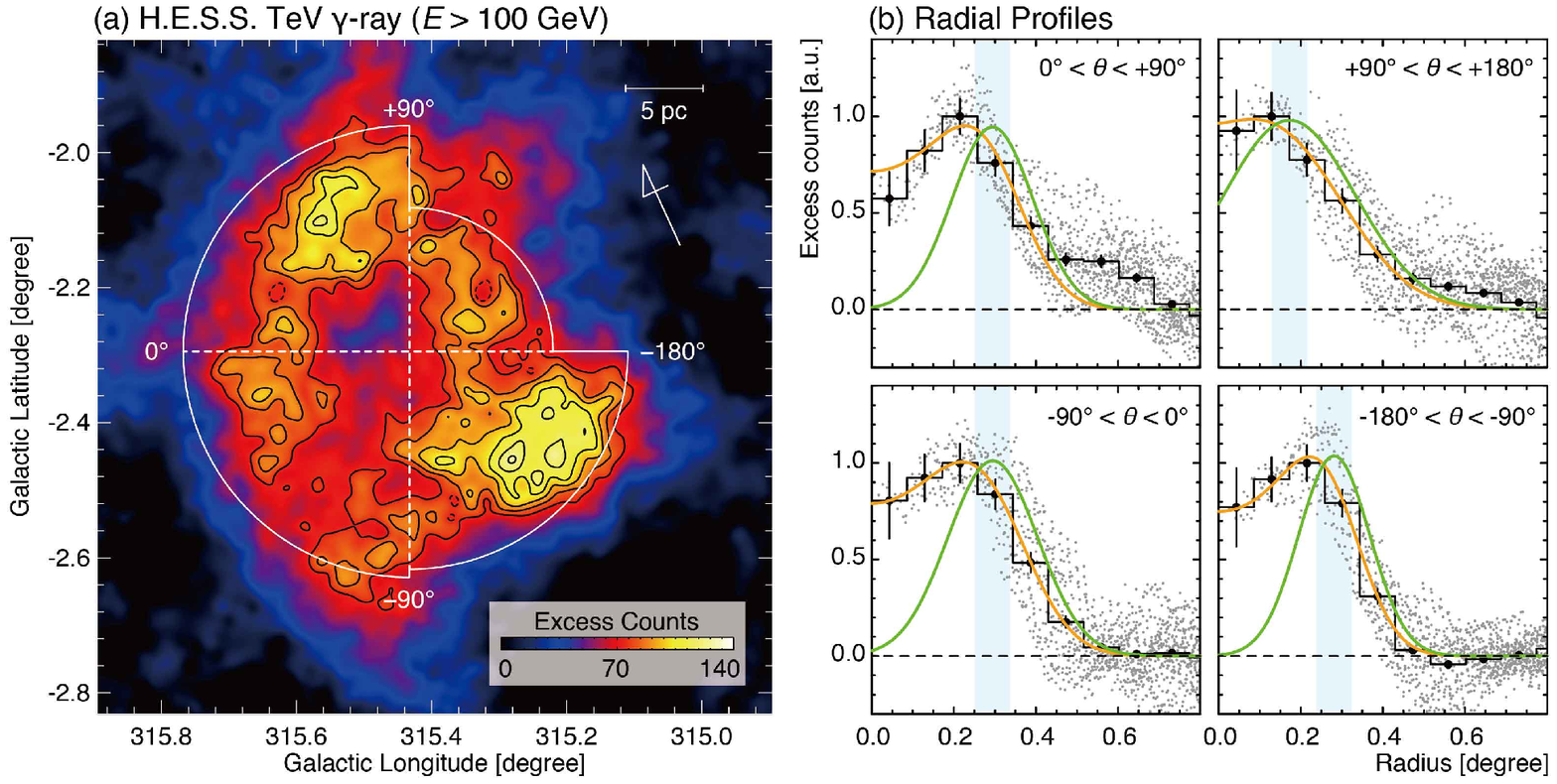}
\caption{(a) Distribution of TeV gamma-ray excess counts of RCW~86 \citep[$E > 100$ GeV, ][]{2018AA...612A...4H}. Black contours correspond to 75, 85, 95, 105, 115, and 125 excess counts. (b) Radial profiles of TeV gamma rays for the four sectors defined in Figure \ref{fig1}a, centered at ($\alpha_\mathrm{J2000}$, $\delta_\mathrm{J2000}$) = ($14^{\mathrm{h}}43^{\mathrm{m}}2\fs16$, $-62{^\circ}$26$\arcmin$56$\arcsec$) or ($l$, $b$) = (315\fdg43, $-2\fdg29$) \citep[see][]{2018AA...612A...4H}. Small dots represent the distributions of all the data points for TeV gamma ray and large filled circles with error bars represent averaged values at each annulus. We adopt a three-dimensional spherical-shell with a Gaussian intensity distribution along its radius to interpolate the TeV gamma-ray distribution (see the text). The green line represents the estimated three-dimensional Gaussian distribution and the orange line represents its projected distribution. Areas shaded in blue correspond to the shell radius for each sector.}
\label{fig1}
\end{center}
\end{figure*}%

\section{Results}\label{sec:results}
\subsection{TeV Gamma-Ray Distribution}\label{subsec:distribution}
Figure \ref{fig1}(a) shows the TeV gamma-ray distribution of RCW~86. Two bright gamma-ray peaks appear toward the north-east at ($l$, $b$) $\sim$ (315\fdg56, $-2\fdg10$) and toward the southwest at ($l$, $b$) $\sim$ (315\fdg24, $-2\fdg44$). A shell-like morphology appears clearly with elongated ellipticity in the south-north direction. The shell radius and thickness of the entire SNR are calculated to be $\sim15$ pc and $\sim5$ pc, respectively \citep{2018AA...612A...4H}. Despite these results, we note that the shell radius of the northwestern part appears to be shorter than that of the other parts. To clarify the azimuthal dependence of the shell radius, we assumed a three-dimensional spherical shell with a single Gaussian function $F(r)$ \citep[e.g.,][]{2012ApJ...746...82F}:
\begin{eqnarray}
F(r) =  A \exp[-(r - r_0\bigr)^2/2\sigma ^2]
\label{eq1}
\end{eqnarray}
where $A$ is a normalization factor, $r_0$ is the radius of the shell in unit of degrees, and $\sigma$ is the standard deviation of the Gaussian function in unit of degrees. We assign the central position of the shell to be ($l$, $b$) = (315\fdg43, $-2\fdg29$), as determined by \cite{2018AA...612A...4H}.

\begin{deluxetable}{cccc}[]
\tablewidth{\linewidth}
\tablecaption{Best-fit parameters of the TeV gamma-ray shell}
\tablehead{Azimuthal {angle} & $r_0$ & $\sigma$  & Extent$^{\dagger}$\\ 
(degrees) & (pc) & (pc) & (pc)\\ 
}
\startdata
\phantom{0}$+90\arcdeg < \theta < +180\arcdeg$ & {\phantom{0}$7.5 \pm 4.6$} & {$7.0 \pm 2.9$} & {\phantom{0}{$9.4 \pm 4.6$}}\\
\phantom{0}\phantom{0}$0\arcdeg < \theta < +90\arcdeg$  & $12.8 \pm 0.9$ & $4.3 \pm 1.1$ & {$14.7 \pm 0.9$}\\
\phantom{0}$-90\arcdeg < \theta < 0\arcdeg$\phantom{0}\phantom{0}\phantom{0}\phantom{0} & $12.9 \pm 1.1$ & $4.8 \pm 1.2$ &{$14.7 \pm 1.1$}\\
\phantom{0}$-180\arcdeg < \theta < -90\arcdeg$\phantom{0}\phantom{0} & $12.3 \pm 0.8$ & $3.8 \pm 0.8$ & {$14.1 \pm 0.8$}\\
\enddata
\tablecomments{$^{\dagger}$ Extent of the shell is defined as $r_0 + \mathrm{PSF} / 2$.}
\label{tab2}
\end{deluxetable}

Figure \ref{fig1}(b) shows the radial scatter profiles of TeV gamma-ray excess counts and an average value, which is shown as a step function in radius $r$ at every 0\fdg086 ($\sim3.8$ pc at 2.5 kpc). We divide the shell into four sectors: first quadrant ($+90\arcdeg < \theta < +180\arcdeg$), second quadrant ($0\arcdeg < \theta < +90\arcdeg$), third quadrant ($-90\arcdeg < \theta < 0\arcdeg$), and forth quadrant ($-180\arcdeg < \theta < -90\arcdeg$), where $\theta$ is the azimuthal angle and is measured clockwise, as shown in Figure \ref{fig1}(a). The best-fit parameters are summarized in Table \ref{tab2}. We find that the shell defined as $r_0 +$ PSF/2 extends $\sim10$ pc in the first quadrant and over $\sim15$ pc in the other quadrants. Hereafter, we focus on the TeV gamma-ray and multiwavelength distributions within the extent of the shell.

\begin{figure*}
\begin{center}
\includegraphics[width=\linewidth,clip]{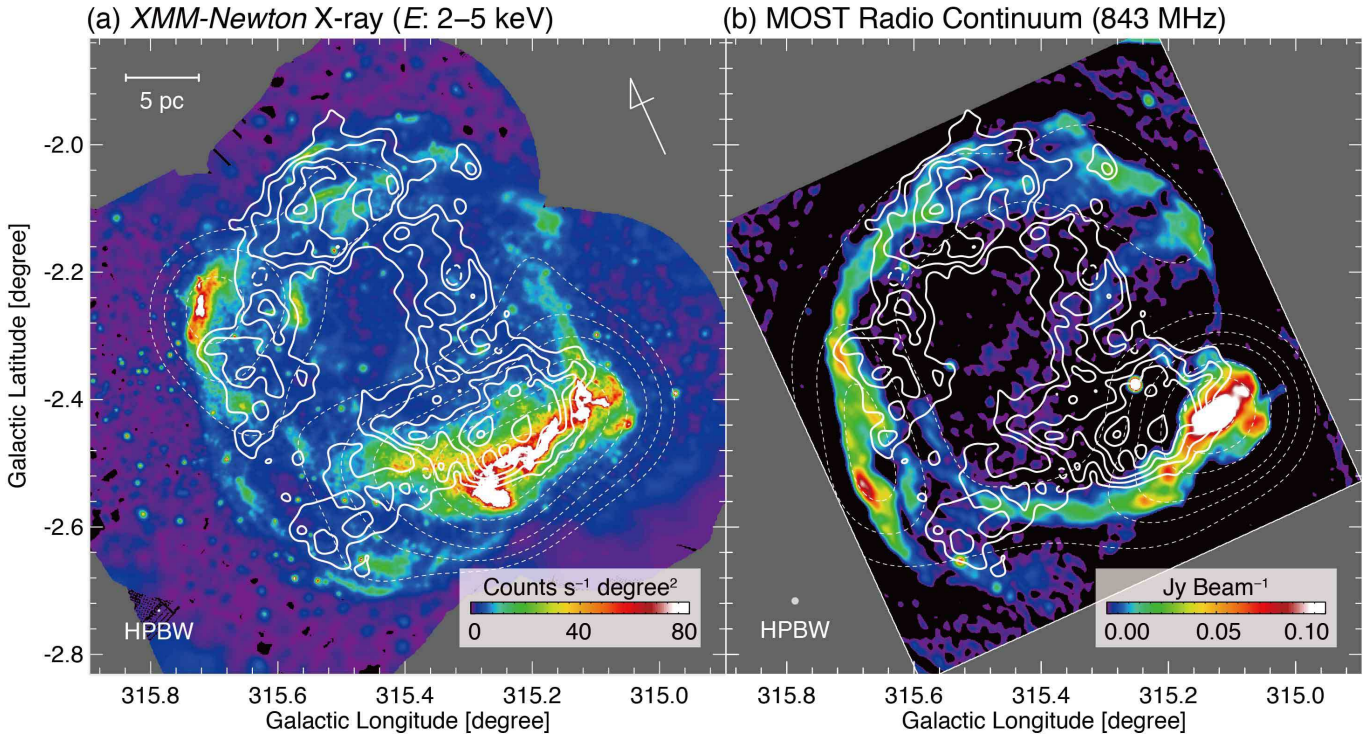}
\caption{Maps of (a) the $XMM$-$Newton$ synchrotron X-ray ($E$: 2--5 keV) and (b) MOST 843 MHz radio continuum. Superposed solid contours indicate the TeV gamma-ray excess counts and contour levels of which are the same as in Figure \ref{fig1}(a). Dashed contours of (a) and (b) indicate the X-ray and radio continuum contours which were smoothed to match the PSF of the TeV gamma-rays. The contour levels are 5,7,10,15, 20, 25, and 30 counts s$^{-1}$ degrees$^{2}$ for (a), and are 0.35, 0.70, 1.05, 1.75, 2.45, and 3.15 mJy Beam$^{-1}$ for (b).}
\label{fig2}
\end{center}
\end{figure*}%

\subsection{Distribution of X-rays and Radio Continuum}\label{sec: synchrotron}
Figures \ref{fig2}(a) and \ref{fig2}(b) show the synchrotron X-rays and radio continuum images of RCW~86 superposed on the TeV gamma-ray contours ($solid$ $lines$). Dashed contours of Figures \ref{fig2}(a) and \ref{fig2}(b) indicate the X-ray and radio continuum contours which were smoothed to match the PSF of the TeV gamma-rays. Both the X-rays and radio continuum show a nearly circularly symmetric shell with no strong emission inside. The most prominent X-ray peak is in the direction of the southwestern shell around ($l$, $b$) $\sim$ (315\fdg2, $-2\fdg5$), which is also bright in TeV gamma rays (see the solid contours in Figures \ref{fig2}(a) and \ref{fig2}(b)). The second brightest X-ray peak at ($l$, $b$) $\sim$ (315\fdg72, $-2\fdg25$) appears to be spatially shifted with respect to the gamma-ray peak, whereas the northern part of the shell shows a good spatial correlation between the X-ray and gamma-ray images. In contrast, the northwestern gamma-ray shell around ($l$, $b$) $\sim$ (315\fdg3, $-2\fdg3$) has no clear counterpart in either the synchrotron X-rays or the radio continuum. This region seems to be located not only inside the shell of the synchrotron X-rays, but also in that of the radio continuum, even after both the X-rays and radio continuum maps were smoothed to match the PSF of the TeV gamma-rays. Moreover, the bright gamma-ray peaks mainly come from inside the radio continuum shell, whereas no counterparts exist in the diffuse and bright radio continuum except in the northern shell.

\begin{figure}
\begin{center}
\includegraphics[width=78mm,clip]{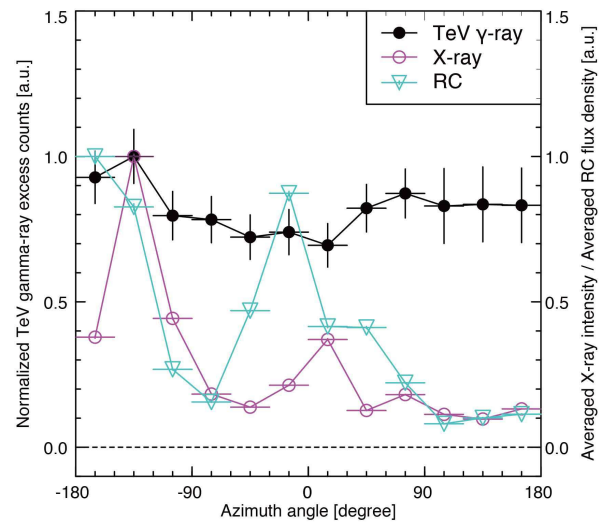}
\vspace*{-0.09cm}
\caption{Azimuthal profiles of TeV gamma-ray excess counts (black filled circles), averaged synchrotron X-ray intensity (magenta open circles) and averaged radio continuum flux density (RC; cyan open triangles) within the region indicated by white solid line in Figure \ref{fig1}(a). Radius of the regions are $\sim0\fdg33$ for the azimuth angle form $-180{^\circ}$ to $+90{^\circ}$ and $\sim0\fdg22$ for the azimuth angle form $+90{^\circ}$ to $+180{^\circ}$.}
\label{fig3}
\end{center}
\vspace*{-0.1cm}
\end{figure}%

\begin{figure*}
\begin{center}
\includegraphics[width=\linewidth,clip]{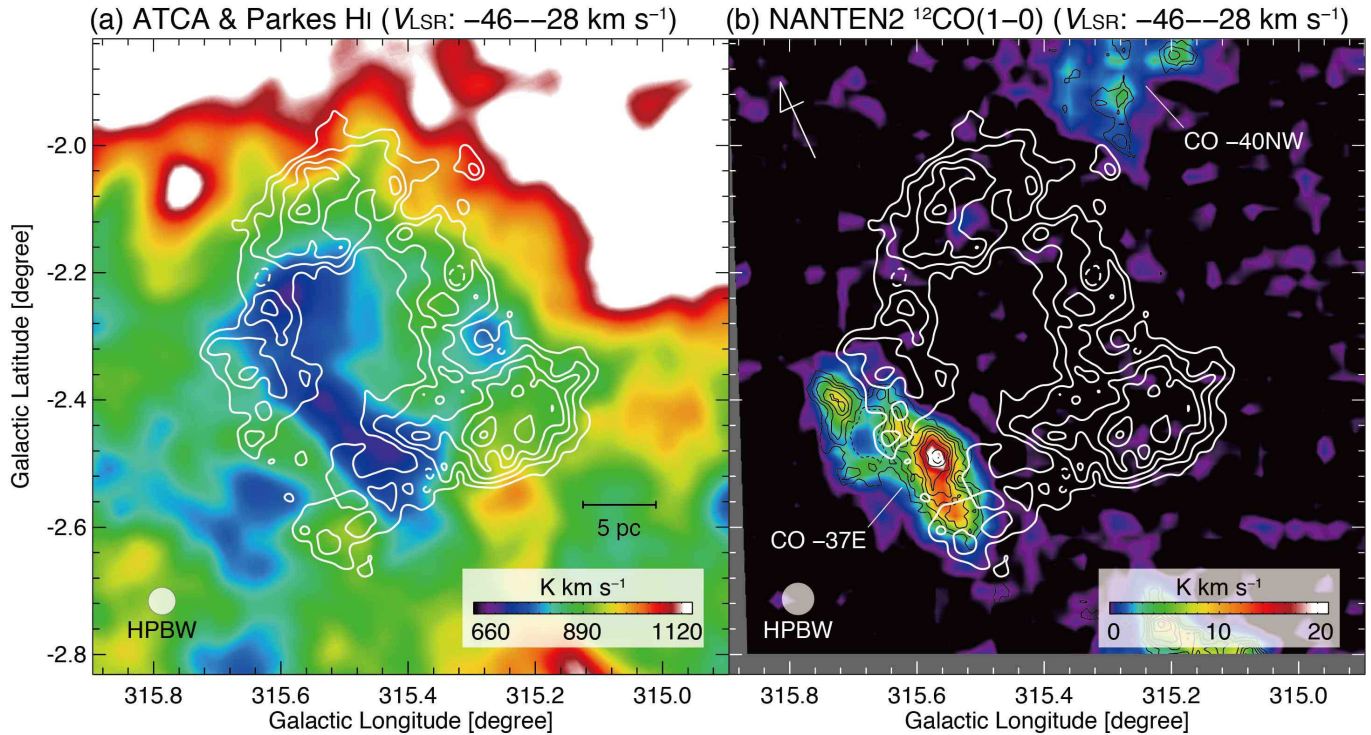}
\caption{Maps of (a) ATCA $\&$ Parkes H{\sc i} and (b) NANTEN2 $^{12}$CO($J$ = 1--0) \citep{2017JHEAp..15....1S}. Both the CO and H{\sc i} are integrated in the velocity range from $-46$ to $-28$ km s$^{-1}$. Superposed white contours indicate the TeV gamma-ray excess counts, with the same contour levels as in Figure \ref{fig1}(a). Also, NANTEN2 $^{12}$CO($J$ = 2--1) emission (black contours) is shown (contour levels of 4, 6, 8, 10, 12, 14 and 16 K km s$^{-1}$).}
\label{fig4}
\end{center}
\end{figure*}%

Figure \ref{fig3} compares the spatial distribution of the synchrotron and TeV gamma-ray radiation as a function of the azimuthal angle defined in Figure \ref{fig1}(a), where the vertical scale is normalized so that the maximum values for each emission coincide. Hereafter, all datasets are averaged every 30$\arcdeg$ in the azimuthal angle within the region indicated by the white solid line in Figure \ref{fig1}(a). {The gamma-rays show a roughly flat distribution with azimuth angle.} The azimuthal distributions of the synchrotron X-rays and radio continuum radiation have {roughly similar trends to each other, but positions of their intensity peaks are offset. There are two intensity peaks of $-135\arcdeg$ and $15\arcdeg$ for the X-rays; $-165\arcdeg$ and $-15\arcdeg$ for the radio continuum,} and no bright emission from $90\arcdeg$ to $180\arcdeg$.
The linear Pearson correlation coefficient (hereafter correlation coefficient) is {$\sim$0.59 for the X-rays and radio continuum,} $\sim0.28$ for the gamma rays and X-rays, and $\sim-0.02$ for the gamma rays and radio continuum. {The spatial difference between the synchrotron X-rays and radio continuum radiation is consistent with the previous studies \citep{2016ApJ...819...98A,2018AA...612A...4H}.}

\subsection{Distribution of H{\sc i} and CO}\label{sec: hi}
Figures \ref{fig4}(a) and \ref{fig4}(b) show the integrated H{\sc i} and CO intensity maps in the velocity range from $V_\mathrm{LSR}$ = $-46$ to $-28$ km s$^{-1}$. The H{\sc i} cavity with an expanding velocity of $\sim8$ km s$^{-1}$ is likely associated with the SNR, as proposed in Paper I. We find that the H{\sc i} distribution with an intensity of $\sim870$--900 K km s$^{-1}$ (green-colored area) spatially coincides with the bright gamma-ray shell except at the regions ($l$, $b$) $\sim$ (315\fdg60, $-2\fdg25$) and (315\fdg40, $-2\fdg55$). No bright gamma rays were detected in the northwestern region where the H{\sc i} intensity is 1000 K km s$^{-1}$ or higher (yellow- and red-colored areas). The most prominent molecular cloud ``CO $-37$ E'' shows a good spatial anti-correlation with the eastern half of the gamma-ray shell. In the northwest, a diffuse molecular cloud named ``CO $-40$ NW'' is located outside the gamma-ray shell. In addition, these molecular clouds show clumpy structures on few parsec scales (see the black contours in Figure \ref{fig4}b).

\subsection{Total Interstellar Protons}\label{sec: ism}
To obtain the total proton column density, we estimate proton column densities in both the molecular and atomic components \citep[e.g.,][]{2012ApJ...746...82F,2017ApJ...850...71F,2013ApJ...768..179Y,2014ApJ...788...94F,2018ApJ...864..161K}. The proton column density of the molecular component $N_\mathrm{p}$(H$_2$) can be derived from the following relation between the molecular hydrogen column density $N$(H$_2$) and the $^{12}$CO($J$ = 1--0) integrated intensity $W$(CO): 
\begin{eqnarray}
N(\mathrm{H_2}) = X \cdot W(\mathrm{CO}) \; \mathrm{(cm^{-2})},\\
N_\mathrm{p}(\mathrm{H_2}) = 2 \times N(\mathrm{H_2}) \; \mathrm{(cm^{-2})}.
\label{eq23}
\end{eqnarray}
where $X$ is a conversion factor between the $N(\mathrm{H_2})$ and $W$(CO). We utilize $X$ = $0.5 \times 10^{20}$ cm$^{-2}$ (K km s$^{-1}$)$^{-1}$, which is derived in the Appendix. The maximum values of $N_\mathrm{p}(\mathrm{H_2})$ are $\sim 2.3 \times 10^{21}$ cm$^{-2}$ in the CO $-37$ E cloud, and $\sim2.0 \times 10^{21}$ cm$^{-2}$ in the CO $-40$ NW cloud.

When we estimate the proton column density in atomic form $N_\mathrm{p}$(H{\sc i}), the optical depth of H{\sc i} should be considered. According to \cite{2015ApJ...798....6F}, $85\%$ of atomic hydrogen in the local interstellar atomic gas is optically thick with respect to the H{\sc i} 21 cm emission (optical depth $\sim0.5$--3), so we cannot use the usual equation assuming optically thin case (optical depth $\ll 1$) as per \cite{1990ARA&A..28..215D}:
\begin{eqnarray}
N_\mathrm{p}(\mathrm{H}{\textsc{i}}) = 1.823 \times 10^{18}  \cdot W(\mathrm{H}{\textsc{i}}) \;(\mathrm{cm}^{-2}),  
\label{eq4}
\end{eqnarray}
where $W$(H{\sc i}) is the integrated intensity of H{\sc i} in units of K km s$^{-1}$. More recently, \cite{2017ApJ...850...71F} derived the optical depth-corrected $N_\mathrm{p}$(H{\sc i}) as a function of $W$(H{\sc i}) \citep[see Figure 9 and Section 4.4 of][]{2017ApJ...850...71F} by using the dust-opacity map at 353 GHz obtained from the $Planck$ and $IRAS$ datasets \citep[for details, see][]{2014A&A...571A..11P} and assuming nonlinear dust property \citep[see][]{2013ApJ...763...55R,2017ApJ...838..132O}. In the present study, we estimate a conversion factor between the optical depth-corrected $N_\mathrm{p}$(H{\sc i}) and $W$(H{\sc i}) using the results of \cite{2017ApJ...850...71F} as a function of $W$(H{\sc i}). We then derive the optical depth-corrected $N_\mathrm{p}$(H{\sc i}) map using the conversion factor and the $W$(H{\sc i}) values of RCW~86. We finally obtain the average optical-depth-corrected $N_\mathrm{p}$(H{\sc i}) to be $\sim$ ($3.8 \pm 0.6) \times 10^{21}$ cm$^{-2}$ in the RCW~86 region, which is $\sim2.3$ times greater than the result in the optically thin case.

\begin{figure*}
\begin{center}
\vspace*{-0.1cm}
\includegraphics[width=\linewidth,clip]{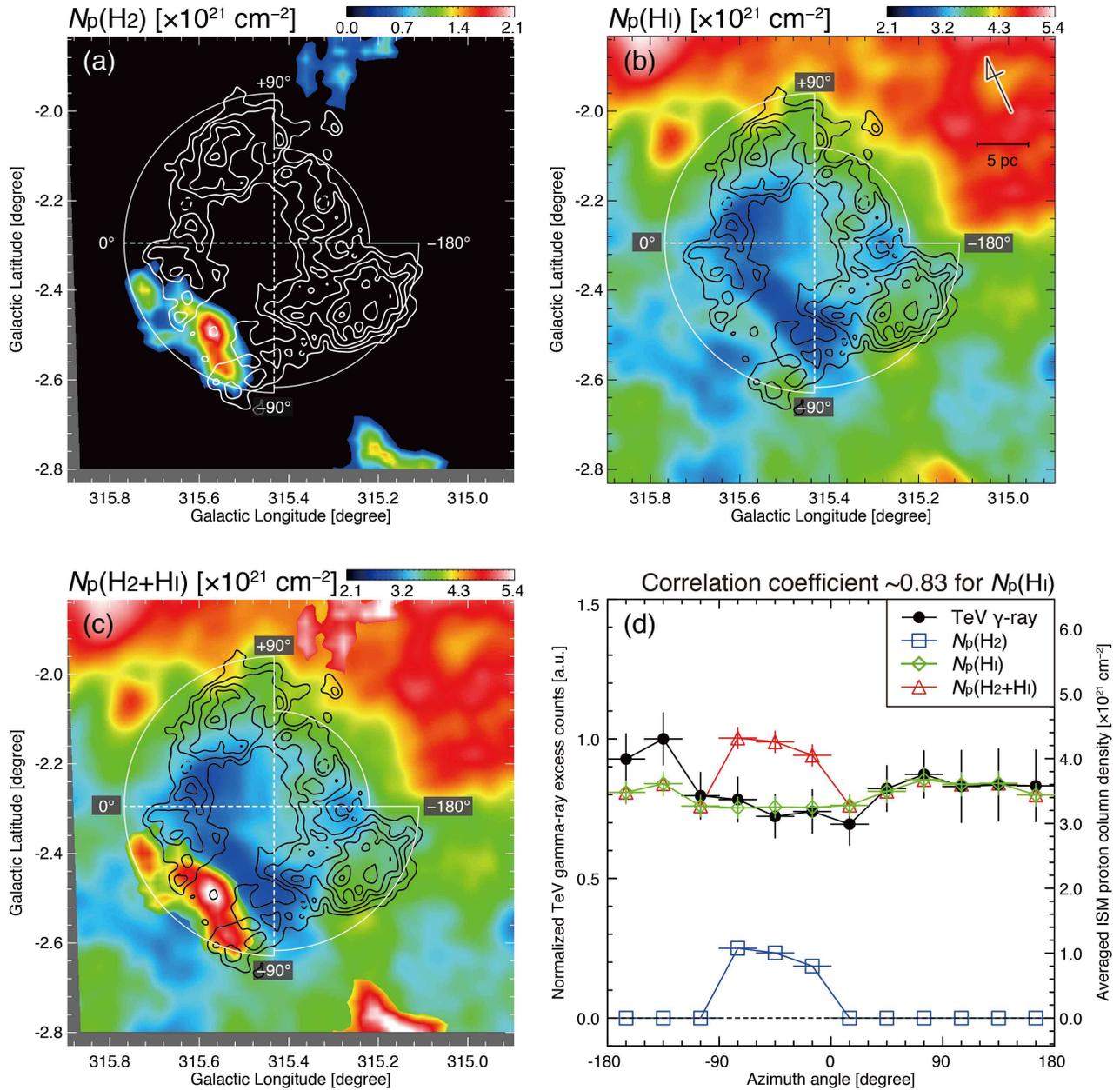}
\caption{Distribution of (a) molecular proton column density $N_\mathrm{p}$(H$_2$), (b) atomic proton column density $N_\mathrm{p}$(H{\sc i}), and (c) total proton column density $N_\mathrm{p}$(H$_2$ + H{\sc i}) in the velocity range from $-46$ to $-28$ km s$^{-1}$. Superposed contours represents for the TeV gamma-ray excess counts, and the contour levels are the same as in Figure \ref{fig1}. (c) Azimuthal profiles of normalized TeV gamma ray, $N_\mathrm{p}$(H$_2$), $N_\mathrm{p}$(H{\sc i}), and $N_\mathrm{p}$(H$_2$ + H{\sc i}), which are averaged every 30$\arcdeg$ in the azimuthal angle within the region indicated by white solid line in Figures \ref{fig5}a, \ref{fig5}b, and \ref{fig5}c.}
\label{fig5}
\end{center}
\end{figure*}%

Figure \ref{fig5}(a), \ref{fig5}(b), and \ref{fig5}(c) show column density maps for the molecular proton $N_\mathrm{p}$(H$_2$), atomic protons corrected for optical depth $N_\mathrm{p}$(H{\sc i}), and the sum of molecular and atomic protons $N_\mathrm{p}$(H$_2 +$ H{\sc i}), respectively. Figure \ref{fig5}(d) shows azimuthal profiles of the normalized TeV gamma rays, $N_\mathrm{p}$(H$_2$), $N_\mathrm{p}$(H{\sc i}), and $N_\mathrm{p}$(H$_2 +$ H{\sc i}). We find that the TeV gamma rays correlate well spatially with $N_\mathrm{p}$(H{\sc i}) except at azimuthal angles from $-180\arcdeg$ to $-120\arcdeg$, where the TeV gamma rays have significant excess counts relative to $N_\mathrm{p}$(H{\sc i}). In contrast, no strong spatial correlations appear between the TeV gamma rays and $N_\mathrm{p}$(H$_2$), while a high correlation with $N_\mathrm{p}$(H$_2 +$ H{\sc i}) is found only between at azimuthal angles from $\sim15\arcdeg$ to $180\arcdeg$. The correlation coefficient is $\sim0.83$ for the gamma rays and $N_\mathrm{p}$(H{\sc i}), $\sim-0.50$ for the gamma-rays and $N_\mathrm{p}$(H$_2$), and $\sim-0.27$ for the gamma rays and $N_\mathrm{p}$(H$_2 +$ H{\sc i}).

\section{Discussion}\label{sec: discussion}
\subsection{Origin of Gamma Rays}\label{sec: origin}
The origin of the gamma rays from RCW~86 has been discussed based on spectral modeling for the gamma ray and synchrotron radiation \citep{2016ApJ...819...98A,2018AA...612A...4H}. However, according to \cite{2012ApJ...744...71I} and \cite{2014MNRAS.445L..70G}, leptonic gamma-rays (inverse Compton scattering of low-energy photon by cosmic-ray electrons) and hadronic gamma-rays (neutral-pion decay from interactions between cosmic rays and interstellar protons) are difficult to distinguish based on spectral modeling alone because of the many uncertain parameters in the model (e.g., magnetic-field strength, spectral index of gamma rays, gas density, and clumpiness). In this section, we discuss that the TeV gamma rays from RCW~86 are likely dominated by the hadronic origin based on spatial comparisons among the gamma-ray, synchrotron radiation, and interstellar gas distribution.

\subsubsection{Morphological Studies}
We first consider a spatial comparison between the gamma-ray and synchrotron radiation. The similarity between the morphology of the gamma-rays and synchrotron X-rays on the 10 pc scale supports the idea that both the gamma rays and X-rays are produced by the same relativistic particles: namely, cosmic-ray electrons \citep[e.g.,][]{2006A&A...449..223A}. For RCW~86, {\cite{2018AA...612A...4H} demonstrated good spatial correspondence between the TeV gamma-rays and synchrotron X-rays using a one-dimensional radial profile toward a specific region, }
which is not inconsistent with a leptonic origin of the TeV gamma rays. {The authors also mentioned that the synchrotron radio and thermal X-ray ($E$: 0.5--1.0 keV) emission regions are distant from the center and dim in the central region of the SNR.} 

{In our detailed comparison using two-dimensional images and azimuthal profiles}, however, we could not find good spatial correspondence between the TeV gamma-rays and synchrotron radiation for the whole SNR (see Figure \ref{fig3}). Moreover, no counterparts of the northwestern TeV gamma-ray shell are found in either the synchrotron X-ray or radio continuum radiation (see also azimuthal angles from 90$\arcdeg$--180$\arcdeg$ in Figure \ref{fig3}). The trend indicates that the ratio between the synchrotron emission power and the inverse Compton effect power, $P_\mathrm{Synch}$/$P_\mathrm{IC}$, is different for each region assuming the pure leptonic model. Since the ratio $P_\mathrm{Synch} / P_\mathrm{IC}$ is proportional to the ratio of energy densities of magnetic field and photon field, $U_\mathrm{B} / U_\mathrm{photon}$, significant spatial variations of the magnetic field strength or photon field should be also observed to explain the trend by the pure leptonic model. However, no significant spatial variations of the magnetic field strength and photon field have been reported due to the limited angular resolution and photon statistics of gamma-rays \citep[e.g.,][]{2016ApJ...819...98A}.

{An alternative idea is that there are two different populations of cosmic-ray electrons which emit the gamma-ray, X-rays, and radio continuum. According to previous X-ray studies, large spatial variations of the physical condition are reported in RCW~86 \citep{2006ApJ...648L..33V,2014MNRAS.441.3040B,2017ApJ...835...34T}. In fact, \cite{2016ApJ...819...98A} demonstrated that the broadband spectral is well fitted by the leptonic scenario using the two-zone model. Furthermore, several publications reported that reverse shock acceleration occurred in the southwestern region where the dense interstellar gas is associated with the SNR \citep{2002ApJ...581.1116R,2016ApJ...819...98A,2017JHEAp..15....1S,2018AA...612A...4H}. In the present gamma-ray dataset, we could not discuss the reverse shock acceleration due to the large PSF. Further spatially resolved spectral modeling and comparative studies with the ISM are needed for testing the scenario.}

Next, we focus on {a spatial comparison between the gamma-ray and interstellar gas.} Assuming an azimuthally isotropic distribution of cosmic rays, {distributions of the leptonic gamma-rays will be reflected that of accelerated cosmic-ray electrons because low-energy seed photons such as cosmic microwave background are uniformly distributed in a scale of SNRs. On the other hand,} the hadronic gamma-ray flux is proportional to the target interstellar gas density. The spatial correspondence between interstellar protons and gamma rays provides one necessary condition for the hadronic origin of gamma rays and can be used decisively to understand the origin of gamma rays {\citep[][]{2003PASJ...55L..61F,2012ApJ...746...82F,2017ApJ...850...71F,2013ASSP...34..249F,1994A&A...285..645A,2006A&A...449..223A,2008A&A...481..401A,2012PASJ...64....8H,2013ApJ...768..179Y,2013MNRAS.434.2188M,2013PASA...30...55M,2018MNRAS.474..662M,2018ApJ...866...76M,2018MNRAS.480..134M,2014ApJ...788...94F,2017MNRAS.464.3757L,2019MNRAS.483.3659L,2017MNRAS.468.2093D,2017ApJ...843...61S,2018ApJ...864..161K}.}
Estimating the total amount of interstellar protons in both atomic and molecular forms is therefore crucial for testing this interpretation \citep[e.g.,][]{2012ApJ...746...82F}.

We argue that the good spatial correspondence between the interstellar proton column density of atomic form, $N_\mathrm{p}$(H{\sc i}), and the TeV gamma rays is possible evidence for hadronically produced gamma rays from RCW~86. Figure \ref{fig5}(a) shows clearly that the TeV gamma-ray shell coincides with an interstellar cavity of atomic hydrogen. In the azimuthal profiles, the TeV gamma rays show almost the same trends as the $N_\mathrm{p}$(H{\sc i}) with a correlation coefficient $\sim0.84$. This result is similar to that of previous studies; e.g., the correlation coefficient between the total interstellar protons and TeV gamma rays is $\sim0.78$ for RX~J1713.7$-$3946 \citep{2012ApJ...746...82F} and $\sim0.95$ for Vela Jr \citep{2017ApJ...850...71F}.

However, to conclude that the gamma rays from RCW~86 are dominantly of hadronic origin, {three} problems should be solved: The first problem is the poor spatial correspondence between the TeV gamma rays and the interstellar proton column density of molecular form, $N_\mathrm{p}$(H$_2$). The second problem is {the very hard {\it Fermi}-LAT spectrum of $\sim$1.42 \citep{2016ApJ...819...98A}.} {The third problem is} the TeV gamma-ray excess relative to $N_\mathrm{p}$(H{\sc i}) at azimuthal angles from $-180\arcdeg$ to $-120\arcdeg$.

\subsubsection{Diffusion Length of Cosmic Rays}
The poor spatial correlation between $N_\mathrm{p}$(H$_2$) and TeV gamma rays can be understood by noting that the low-energy cosmic rays do not penetrate into dense molecular clouds. According to \cite{2012ApJ...744...71I}, the penetration depth $l_\mathrm{pd}$ of cosmic rays is 
\begin{eqnarray}
l_\mathrm{pd} =  0.1\; \eta^{0.5}\;  (E / 10\;\mathrm{TeV})^{0.5}\; (B / 100\;\mathrm{\mu G})^{-0.5}\nonumber \\
(t_\mathrm{age} / 1000\;\mathrm{yr})^{0.5} \;\;(\mathrm{pc}),
\label{eq5}
\end{eqnarray}
where $\eta$ is a turbulence-factor defined as the degree of magnetic-field fluctuations ($\eta$ $= B^2/\delta B^2 \ga 1$), $E$ is the cosmic-ray energy, $B$ is the magnetic-field strength, and $t_\mathrm{age}$ is the age of the SNR. In RCW~86, a TeV gamma-ray image is shown for energies above $\sim100$ GeV, which means that the image traces the spatial distribution of cosmic rays with energies of $\sim1$ TeV {or higher} if the hadronic process dominates. {Since the TeV gamm-ray spectrum is steep ($\Gamma = -2.3$) with a low cutoff ($\sim3.5$ TeV), most of the TeV events are from a few TeV \citep{2018AA...612A...4H}.} The age of the SNR is $\sim1800$ yr \citep{1975Obs....95..190C,2006ChJAA...6..635Z}.

\begin{figure*}
\begin{center}
\includegraphics[width=\linewidth,clip]{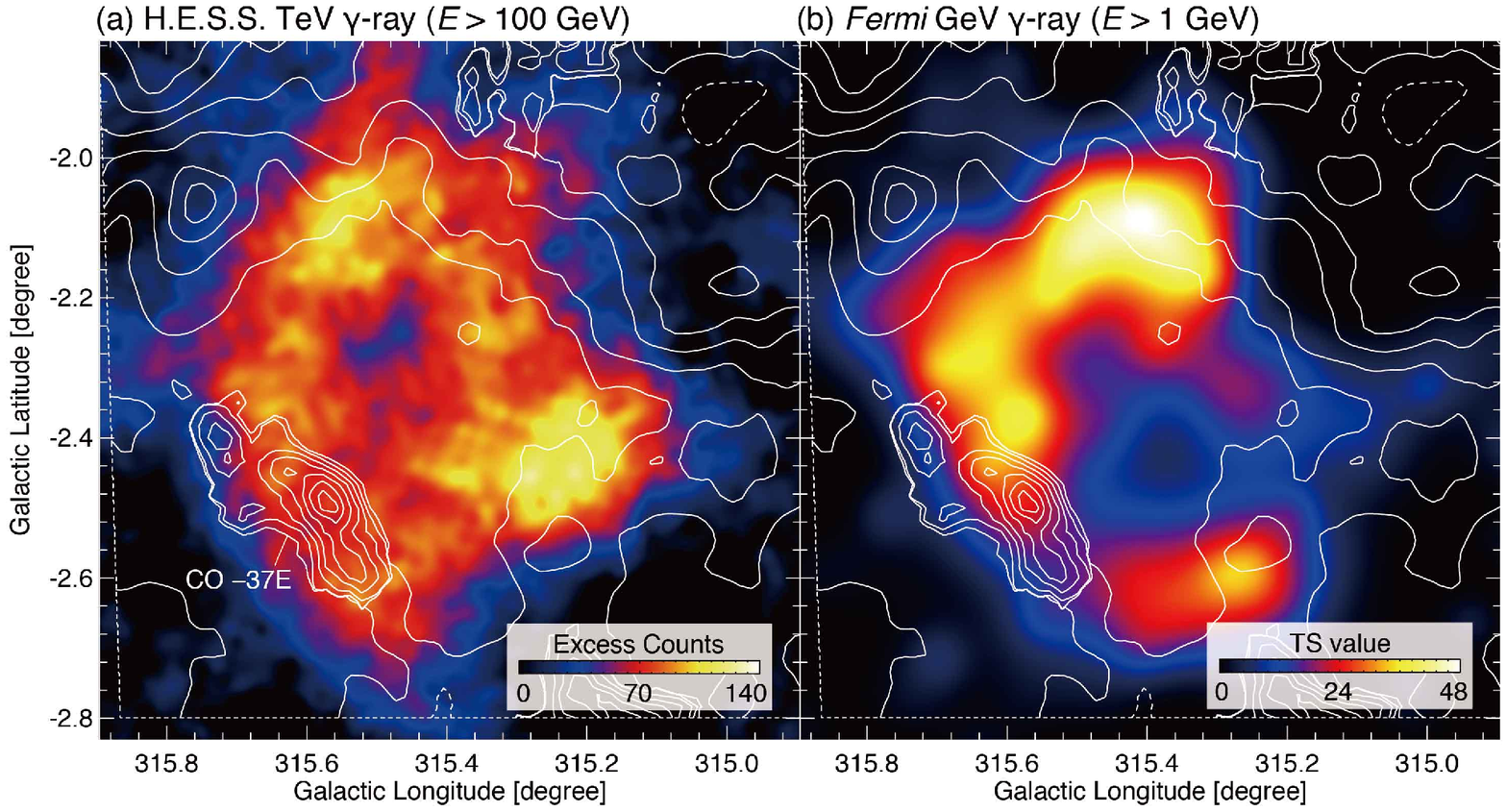}
\caption{{Distribution of (a) H.E.S.S. TeV gamma-rays \citep[$E > 100$ GeV, ][]{2018AA...612A...4H} and (b) {\it Fermi} LAT GeV gamma-rays  \citep[$E > 1$ GeV, ][]{2016ApJ...819...98A}. The superposed contours represent the total proton column density $N_\mathrm{p}$(H$_2$ + H{\sc i}) as shown in Figure \ref{fig5}c. The lowest contour level and contour intervals are $3.6 \times 10^{21}$ cm$^{-2}$ and $0.3 \times 10^{21}$ cm$^{-2}$, respectively.}}
\label{fig6}
\vspace*{0.3cm}
\end{center}
\end{figure*}%

The magnetic field strength and turbulence-factor $\eta$ ($= B^2/\delta B^2$) both have large ambiguity. {The} magnetic-field strength $B$ in {interstellar quiescent gas clouds} is given by the following equation \citep{2010ApJ...725..466C}:
\begin{eqnarray}
B \sim 10\;(n / 300\; \mathrm{cm^{-3}})^{0.65} \;\; \mathrm{(\mu G)}
\label{eq6}
\end{eqnarray}
where $n$ is the number density of the interstellar gas. According to Paper I, the CO cloud ``CO $-37$ E'' has a number density of $\sim300$ cm$^{-3}$, which gives a magnetic-field strength of $\sim10$ $\mu$G. {It should be noticed that the value is a lower limit because the magnetic-field strength is measured in quiescent clouds without any shock disturbance. In the case of a shocked region, the strong magnetic field has been reported in the young SNRs \citep[e.g.,][]{2002A&A...396..649V,2003ApJ...584..758V,2004AdSpR..33..376B,2007Natur.449..576U}. In fact,} \cite{2005A&A...433..229V} derived a magnetic-field strength of $\sim100$ $\mu$G from the morphology of synchrotron X-rays in the southwestern filaments of RCW~86. {Furthermore}, Paper I shows that shock-cloud interaction occurs around the molecular clouds associated with RCW~86. When the SNR shock waves interact with the clumpy interstellar gas, the turbulence and magnetic field are strongly enhanced around the gas clumps, which are rim-brightened by synchrotron radiation (both X-ray and radio continuum). According to three-dimensional magnetohydrodynamic simulations, the average magnetic field is $\sim30$ $\mu$G downstream of the shock, and the turbulence-factor is $\sim1$, which corresponds to Bohm-limit diffusion \cite[][]{2012ApJ...744...71I}.

Here we therefore use {$E$ = 10 TeV}, $t_\mathrm{age}$ = 1800 yr, $B$ = 10--100 $\mu$G, and $\eta$ = 1, so the penetration depth is $l_\mathrm{pd}$ = {0.13--0.42 pc}. The result is significantly less than the size of the molecular clouds, which are $\sim5$--12 pc (see Paper I). We therefore conclude that the low-energy cosmic rays traced by the TeV gamma-ray image of \cite{2018AA...612A...4H} cannot, at the moment, penetrate into the dense molecular gas. This means that the dense molecular cloud is not a target of the low-energy cosmic rays, and a strong spatial correspondence between the TeV gamma rays and $N_\mathrm{p}$(H$_2$) or $N_\mathrm{p}$(H$_2 +$ H{\sc i}) is not expected.

{The energy-dependent diffusion of cosmic rays is also consistent with the GeV gamma-ray map of RCW~86 obtained with {\it Fermi}-LAT. Figure \ref{fig6} shows TeV gamma-ray excess image as shown in Figure \ref{fig1} \citep[$E > 100$ GeV, ][]{2018AA...612A...4H} and the TS map of {\it Fermi} GeV gamma-rays \citep[$E > 1$ GeV, ][]{2018AA...612A...4H}. The GeV gamma-ray map shows a dip-like structure toward the dense molecular cloud CO $-37$E; the significance level toward the CO cloud ($\sqrt{TS} \sim$4$\sigma$) is 1.5 times smaller than the gamma-rays surrounding the cloud ($\sqrt{TS} \sim$6$\sigma$). On the other hand, the TeV gamma-rays are still bright toward the molecular cloud. The TeV gamma-ray excess toward the CO cloud is only $\sim$1.3 times lower than the TeV gamma-rays surrounding the cloud. Since the GeV gamma-ray map (Figure \ref{fig6}b) traces low-energy cosmic-ray protons with $E \sim 10$ GeV, the penetration depth is expected to be about 10 times shorter than the high-energy cosmic rays with a few TeV as shown by the TeV gamma-ray image (Figure \ref{fig6}a). We therefore conclude that the low-energy cosmic-rays traced by {\it Fermi} GeV gamma-ray map cannot penetrate into the dense molecular cloud at this time. To confirm the scenario, we need further gamma-ray images at the energy above 10~TeV, which can trace the $\sim$100 TeV cosmic-ray protons if the hadronic process dominates. In this case, the bright gamma-ray emission will be found toward the dense molecular cloud in RCW~86.}

{We also argue that} the low-energy cosmic rays {of a few TeV} can penetrate into the H{\sc i} clouds and produce TeV gamma-rays via the hadronic process. \cite{2018ApJ...860...33F} carried out synthetic observations of H{\sc i} utilizing magneto-hydrodynamical simulations of the inhomogeneous turbulent ISM \citep{2012ApJ...744...71I}. The authors revealed that the cold neutral medium (cold/dense H{\sc i}) shows a highly clumpy distribution with a volume filling factor of 3.5$\%$, while the warm neutral medium (low-density ambient H{\sc i}) shows a diffuse distribution with a large filling factor of 96.5$\%$. The typical width of the cold/dense H{\sc i} filaments is $\sim 0.1$ pc, significantly smaller than the size of the molecular cloud associated with RCW~86, but is similar to the penetration depth of cosmic rays.

{Based on the above discussion, the very hard {\it Fermi}-LAT spectrum of $\sim$1.42 in RCW~86 can be also explained by the penetration length of cosmic ray protons. If the GeV gamma-rays taken with {\it Fermi}-LAT are hadronic origin, the energy of accelerated cosmic-ray protons is a few GeV. In this case, the penetration length of cosmic ray protons at 10 GeV is to be $\sim$0.001--0.004 pc, which is significantly smaller than the size of the cold/dense H{\sc i} filaments ($\sim$0.1 pc) as shown by \cite{2018ApJ...860...33F}. Therefore, cosmic-ray protons traced by using {\it Fermi}-LAT cannot penetrate neither into dense molecular clouds nor into the cold/dense H{\sc i} filaments in RCW~86. We therefore {expected a very hard GeV gamma-ray spectrum} because effective target interstellar protons will be reduced for the cosmic-ray protons with an energy of $\sim$1 GeV \citep[see also][]{2012ApJ...744...71I,2014MNRAS.445L..70G}. An alternative idea is due to the strong modified shock. According to \cite{2018AA...612A...4H}, the broad-band gamma-ray spectra of RCW~86 can be well described by the hadron-dominant model with a proton spectral index of $\Gamma_\mathrm{p}\sim$1.7 and a cut-off energy of 35 TeV. The index of $\Gamma_\mathrm{p}\sim$1.7 is different from $\Gamma_\mathrm{p} = 2$ with the standard DSA model, but is expected under the strong modified shocks \citep{1999ApJ...511L..53M,1999ApJ...526..385B}.}

\subsubsection{Contributions of Leptonic Gamma Rays in the Southwestern Shell}
We argue that the leptonic gamma rays may be responsible for the gap between the TeV gamma rays and $N_\mathrm{p}$(H{\sc i}) at azimuthal angles from $-180\arcdeg$ to $-120\arcdeg$. The regions in which TeV gamma-ray excess occurs correspond to the southwestern shell of the SNR, which emits the brightest peaks of both synchrotron X-rays and radio continuum radiation [Figures \ref{fig2}(a) and \ref{fig2}(b)].

A similar situation also occurs for the TeV gamma rays from SNR HESS~J1731$-$347. \cite{2014ApJ...788...94F} found a significant difference (factor of two) in the azimuthal profile between the normalized interstellar proton column density and the gamma rays from the southern part of the shell. Both synchrotron X-rays and radio continuum are bright but show a significantly lower density of $\sim10$ cm$^{-3}$. They suggested that the low-density gas may lead to lower amplification of the magnetic fields via shock-cloud interaction and thereby favor the leptonic origin of the TeV gamma rays in the south of HESS~J1731$-$347, because the synchrotron cooling of cosmic-ray electrons is not significant. {The authors estimated the leptonic contribution to $\sim20$\% of the total gamma rays in HESS~J1731$-$347.}

For RCW~86, the southwestern shell has a large amount of gas ($\sim150$ cm$^{-3}$) in the cloud H{\sc i} clump (see Paper I), and the magnetic field is strongly enhanced \citep[e.g.,][]{2005A&A...433..229V}. The difference between the $N_\mathrm{p}$(H{\sc i}) and the TeV gamma rays is therefore expected to be smaller than for HESS~J1731$-$347. {According to the SED modeling by \cite{2018AA...612A...4H}, the peak flux of the hadronic gamma-ray is $\sim$18 times higher than that of the leptonic gamma-ray in the hadron-dominant model \citep[see Figure 5 in][]{2018AA...612A...4H}. We therefore propose that the leptonic gamma rays accounts for only $\sim$6\% of the total gamma-ray flux in RCW~86, which represents the gap between the TeV gamma rays and $N_\mathrm{p}$(H{\sc i}) at azimuthal angles as shown in Figure \ref{fig5}(d).}


We also note slight differences between $N_\mathrm{p}$(H{\sc i}) and gamma rays at the azimuthal angles of $\sim15\arcdeg$ and $\sim-90\arcdeg$, which correspond to regions with low $N_\mathrm{p}$(H{\sc i}) [see blue-colored area in Figure \ref{fig5}(a)]. These regions have the potential to provide significant contributions from leptonic gamma rays, because of the low magnetic field and gas density. However, we will not discuss it further due to the limited angular resolution of the TeV gamma-ray data. Further gamma-ray observations using the Cherenkov Telescope Array (CTA) should provide a gamma-ray image with high-spatial resolution necessary to allow us to distinguish the differences between $N_\mathrm{p}$(H{\sc i}) and gamma rays.

\subsubsection{{Comparison with RX~J1713.7$-$3946}}
We also discuss about the difference between the SNRs RCW~86 and SNR RX~J1713.7$-$3946. The main physical properties of both SNRs are listed in Table \ref{tab3} along with other TeV SNRs which will be discussed later. Both RCW~86 and RX~J1713.7$-$3946 are quite similar in age, diameter and total proton mass associated. On the other hand, the TeV gamma-ray of RX~J1713.7$-$3946 shows good spatial correspondence with the total interstellar proton column density $N_\mathrm{p}$(H$_2 +$ H{\sc i}) \citep{2012ApJ...746...82F}, while that of RCW~86 spatially coincides with only the interstellar protons in atomic form $N_\mathrm{p}$(H{\sc i}). One should consider the possibility that the size of molecular clouds is slightly different between the SNRs RCW~86 and RX~J1713.7$-$3946. According to \cite{2005ApJ...631..947M}, typical size of molecular clouds associated with RX~J1713.7$-$3946 is $2.8 \pm 0.6$ pc, while that of CO~$-37$~E cloud in RCW~86 is $\sim12.4$ pc (see Paper I). The low-energy cosmic rays in RCW~86 are therefore hard to penetrate into the denser part of molecular clouds owing to its large size.

\begin{deluxetable*}{lcccc}[]
\tablewidth{\linewidth}
\tablecaption{Comparison of physical properties in the young shell-type SNRs}
\tablehead{
& RX~J1713.7$-$3946$^\mathrm{a)}$ & RCW~86$^\mathrm{b)}$ &Vela~Jr$^\mathrm{c)}$ & HESS~J1731$-$347$^\mathrm{d)}$}
\startdata
Age (yr) & 1600 & 1800 & 2400 & 4000\\
Distance (kpc) & 1 & 2.5 & 0.75 & 5.2\\
Radius (pc) & 8.2 & 7.5 & 5.9 & 11\\
Molecular proton mass (10$^4$ $M_{\odot}$) & 0.9 & --- & 0.1 & 5.1\\
Atomic proton mass (10$^4$ $M_{\odot}$) & 1.1 & 2.0 & 2.5 & 1.3\\
Total proton mass (10$^4$ $M_{\odot}$) & 2.0 & 2.0 & 2.6 & 6.4\\
$N_\mathrm{p}$(H$_2$) (cm$^{-3}$) & 60 & --- & 4 & 48\\
$N_\mathrm{p}$(H{\sc i}) (cm$^{-3}$) & 70 & 75 & 96 & 12\\
$N_\mathrm{p}$(H$_2$ + H{\sc i}) (cm$^{-3}$) & 130 & 75 & 100 & 60\\
$N_\mathrm{p}$(H$_2$)/$N_\mathrm{p}$(H{\sc i}) & 0.9 & $< 0.01$ & 0.04 & 4\\ 
Total cosmic-ray energy ($10^{48}$ erg)$^\dagger$ & 0.4 & 1.2 & 0.7 & 7\\ 
SNR Type & CC & Type Ia & CC? & CC\\
\enddata
\label{tab3}
\tablecomments{$^\dagger$ We adopt $N_\mathrm{p}$(H$_2$ + H{\sc i}) {and distances} to the latest hadronic model{s} presented by \cite{2018AA...612A...6H} {for RX~J1713.7$-$3946, \cite{2018AA...612A...4H} for RCW~86, \cite{2018AA...612A...7H} for Vela~Jr., and \cite{2018ApJ...853....2G} for HESS~J1731$-$347}.} 
\end{deluxetable*}

In addition, the shock-interacting time is also possibly different between both SNRs. For RX~J1713.7$-$3946, all molecular clouds associated with the SNR have been impacted by shock waves for the last $1000$ yrs \citep[e.g.,][]{2016scir.book.....S}. In contrast, shock waves of RCW~86 reached CO~$-37$~E cloud very recently. According to \cite{2008PASJ...60S.123Y}, the time elapsed since the Fe-rich ejecta was heated by the reverse shock was estimated to be $\sim 380$ yr or less. Although the molecular cloud appears to be embedded within the SNR in the projected image (see Paper I), the short time elapsed can be explained in terms of the shock-cloud interaction with an inclination angle to the line of sight. If CO~$-37$~E cloud is hit by shock waves from the direction perpendicular to the line of sight, the photometric absorption will be observed \citep[e.g.,][]{2015ApJ...799..175S,2016scir.book.....S}. However, there is no shadowing effect due to the molecular cloud in the soft-band X-ray image (0.5--1.0 keV, see Paper I). This trend is also consistent with the X-ray spectroscopy using $Suzaku$ datasets. The absorbing column density of the southeastern shell is $\sim (2.1 \pm 0.1) \times 10^{21}$ cm$^{-2}$, which is the smallest value of the whole SNR \citep{2017ApJ...835...34T}. We finally conclude that the CO~$-37$~E cloud lies behind the SNR and the shock has only recently reached it, hence cosmic rays accelerated in the shock waves could not produce the bright gamma-rays at this time.

An alternative idea is that the density profiles of the molecular clouds are different. \cite{2010ApJ...724...59S} revealed that the molecular core C embedded in RX~J1713.7$-$3946 has a density gradient following a $r^{-2.2\pm0.4}$ law, where $r$ is the radius of the molecular core. If the density gradient of the CO~$-37$~E cloud in RCW~86 is steeper than that of core C in RX~J1713.7$-$3946, the mass illuminated {by cosmic rays from RCW~86 should be further reduced} than that of RX~J1713.7$-$3946. To test this interpretation, we need additional observations of the CO $-37$ E cloud in multiple CO transitions using ASTE, Mopra, and ALMA.

\subsection{Total Cosmic-Ray Energy Budget}
As discussed in Section \ref{sec: origin}, the TeV gamma rays from RCW~86 favor the hadron-dominant origin. To obtain the total cosmic-ray energy, we first derive the number density of interstellar protons associated with the SNR. By using the result of Figure \ref{fig5}(a), the average number density and mass of interstellar atomic protons are estimated to be $\sim75$ cm$^{-3}$ and $\sim2 \times 10^4$ $M_{\sun}$, respectively, assuming a shell radius of $\sim15$ pc and a thickness of $\sim5$ pc \citep{2018AA...612A...4H}. The average number density is roughly consistent with the estimate made in Paper I. The total cosmic-ray energy $W_\mathrm{p}$ above 1 GeV can be derived by using {the hadron-dominant model in} \cite{2018AA...612A...4H}
\begin{eqnarray}
W_\mathrm{p} \sim 9 \times 10^{49} (n / 1\; \mathrm{cm^{-3}})^{-1} \;\; \mathrm{(erg)}, 
\label{eq7}
\end{eqnarray}
where $n$ is the number density of interstellar protons associated with the SNR. Adopting $n$ = 75 cm$^{-3}$, we finally obtain $W_\mathrm{p} \sim1.2 \times 10^{48}$ erg. Table \ref{tab3} compares the young TeV gamma-ray SNRs---RX~J1713.7$-$3946, Vela~Jr, HESS~J1731$-$347, and RCW~86. The total cosmic-ray energy is roughly similar for these SNRs, indicating that both core-collapse and Type Ia supernovae can accelerate cosmic rays up to a total energy of $\sim10^{48}$--$10^{49}$ erg, corresponding to $0.1\%$--$1\%$ of the typical kinematic energy of a supernova explosion ($\sim10^{51}$ erg). These values are significantly below the $\sim10\%$ ($\sim 10^{50}$ erg) derived from conventional expections \citep[e.g.,][]{2008AIPC.1085..265G}. The difference is mainly due to an effect of the filling factor of the interstellar gas within the shell. Generally, the interstellar gas is not uniformly distributed within the shell of SNRs \citep[e.g.,][]{2003PASJ...55L..61F,2018ApJ...864..161K}. Additionally, recent numerical studies of H{\sc i} support the presence of small-scale clumpy structures of cold/dense H{\sc i} \citep[e.g.,][]{2018ApJ...860...33F}. We can only detect hadronic gamma rays in the regions where the interstellar gas is located. Although it is very difficult to derive the filling factor of interstellar gas within the shell, our estimation can be considered a lower limit of the total cosmic-ray energy. {Future GASKAP H{\sc i} observations using the Australian Square Kilometre Array Pathfinder \citep[ASKAP;][]{2009IEEEP..97.1507D} will spatially resolve the small-scale structure of atomic hydrogen with a high-angular resolution of $\sim$30 arcsec.}

\begin{figure*}
\begin{center}
\includegraphics[width=168mm,clip]{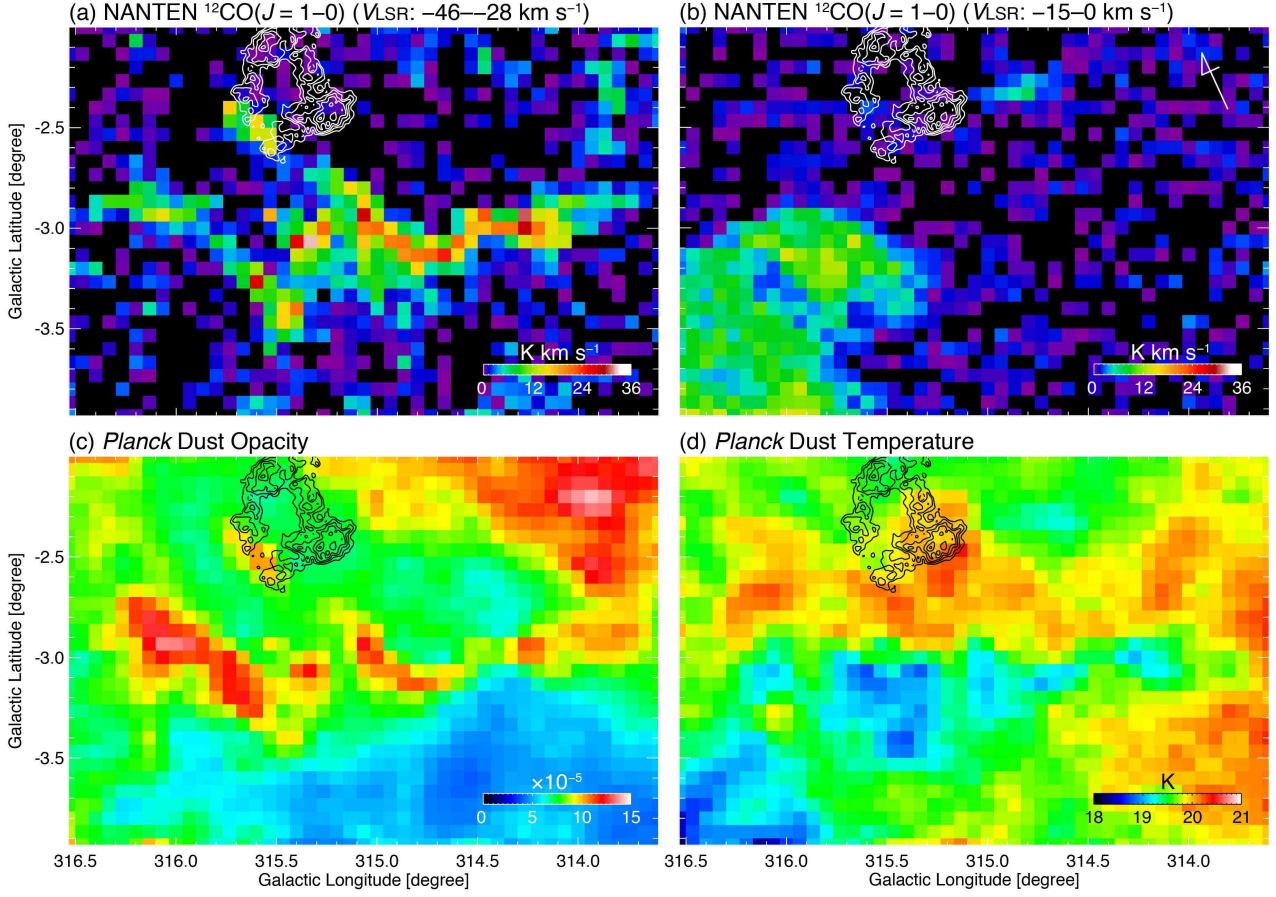}
\caption{Distribution of (a, b) NANTEN $^{12}$CO($J$ = 1--0), (c) $Planck$ dust opacity $\tau_{353}$ at the frequency of 353 GHz, and (c) $Planck$ dust temperature toward RCW~86. The integration velocity ranges are (a) $V_\mathrm{LSR}$ = $-46$--$-28$ km s$^{-1}$ and (b) $V_\mathrm{LSR}$ = $-15$--0 km s$^{-1}$. Superposed contours are the same as Figure \ref{fig1}.}
\label{figa1}
\end{center}
\end{figure*}%

\begin{figure}
\begin{center}
\includegraphics[width=74mm,clip]{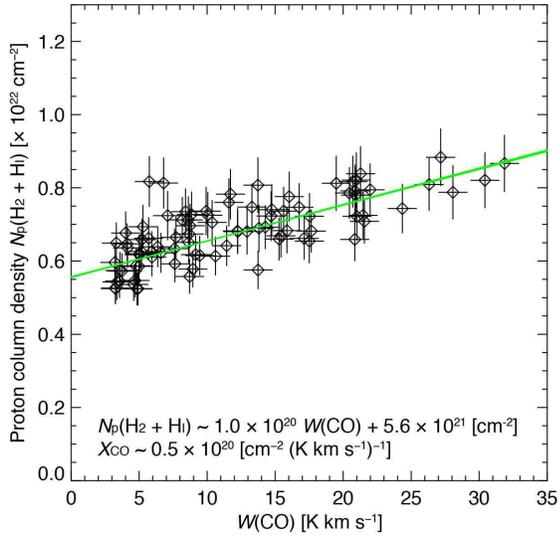}
\caption{Correlation plot between $W$(CO) and $N_\mathrm{H}$ derived using the dust opacity $\tau_{353}$ and equation \ref{eq7}. The linear regression by $\chi^2$ fitting is shown by the solid green line.}
\label{figa2}
\end{center}
\end{figure}%

\section{Conclusions}\label{sec: conclusions}
We present herein a detailed morphological study of TeV gamma rays, synchrotron radiation, and interstellar gas to investigate whether the TeV gamma rays from RCW~86 are dominantly originating from leptonic or hadronic mechanisms. The main conclusions are as follows:

\begin{enumerate}
\item The proton column density of the atomic gas shows good spatial correlation with the TeV gamma rays in the azimuthal profiles, indicating that the TeV gamma rays from RCW~86 are likely dominated by a hadronic origin. In contrast, poor spatial correlation is seen in the southeastern shell of the SNR between the proton column densities of molecular gas and the TeV gamma rays. We argue that this poor correlation may be understood by considering that the low-energy cosmic rays ($\sim1$ TeV) do not penetrate into dense molecular clouds because of the young SNR age, $\sim 1800$ yr, an enhancement of the turbulent magnetic field around the dense cloud of $\sim10$--100 $\mu$G, and a turbulence-factor of $\sim1$.
\item The southwestern shell shows a significant gamma-ray excess that does not reflect the distribution of the proton column density in atomic form. We argue that the leptonic gamma rays possibly contribute as well as the hadronic gamma rays, which is consistent with the enhancement of both the synchrotron X-rays and radio continuum from the southwestern shell. If this is the case, the TeV gamma rays from RCW~86 would mainly consist of hadron-dominant gamma rays over the entire SNR, but the contribution of the leptonic gamma rays to total flux of gamma ray is {$\sim$6\%}.
\item The total cosmic-ray energy is roughly similar for all the young TeV gamma-ray SNRs RX~J1713.7$-$3946, Vela~Jr, HESS~J1731$-$347 and RCW~86, indicating that cosmic rays can be accelerated by both core-collapse and Type Ia supernovae. The total energy of cosmic-rays, $\sim10^{48}$--10$^{49}$ erg, derived by using the gas density gives a safe lower limit, mainly because of the low filling factor of interstellar gas within the shell. 
\end{enumerate}

\acknowledgments
We acknowledge Anne Green for her valuable support during the H{\sc i} observations and reduction, and Gloria M. Dubner, PI of the ATCA project C1011 carried out to obtain the reported H{\sc i} data, who provided them to Yasuo Fukui. {We also acknowledge Marianne Lemoine-Goumard for sending us the GeV gamma-ray dataset obtained with {\it Fermi}-LAT.} This study was based on observations obtained with $XMM$-$Newton$, an ESA science mission with instruments and contributions directly funded by ESA Member States and NASA. We also utilize data from the Molonglo Observatory Synthesis Telescope (MOST), which is operated by The University of Sydney with support from the Australian Research Council and the Science Foundation for Physics within The University of Sydney. The Australia Telescope Compact Array (ATCA) is part of the Australia Telescope National Facility which is funded by the Australian Government for operation as a National Facility managed by CSIRO. NANTEN2 is an international collaboration of 11 universities: Nagoya University, Osaka Prefecture University, University of Bonn, University of Cologne, Seoul National University, University of Chile, University of Adelaide, University of New South Wales, Macquarie University, University of Sydney, and University of ETH Zurich. This study was financially supported by Grants-in-Aid for Scientific Research (KAKENHI) of the Japanese Society for the Promotion of Science (JSPS, grant Nos. 12J10082, 24224005, 15H05694, and 16K17664). H.S. was supported by ``Building of Consortia for the Development of Human Resources in Science and Technology'' of Ministry of Education, Culture, Sports, Science and Technology (MEXT, grant No. 01-M1-0305). E.M.R. is member of the Carrera del Investigador Cient\'\i fico of CONICET (Argentina), and is partially supported by CONICET grant{s} PIP 112-201207-00226 {and 112-201701-00604}. {We are grateful to the anonymous referee for useful comments, which helped the authors to improve the paper.}
\software{ESAS \citep{2008A&A...478..575K}}

\section*{Appendix\\Determination of conversion factor}\label{sec: appendix}
To derive the conversion factor $X$ between the molecular hydrogen column density $N$(H$_2$) and integrated intensity of CO, $W$(CO), we utilize the $Planck$ dust opacity $\tau_{353}$ at the frequency of 353 GHz, dust temperature, and NANTEN $^{12}$CO($J$ = 1--0) maps following the method which is presented by \cite{2017ApJ...838..132O}.

Figures \ref{figa1}a and \ref{figa1}b show the NANTEN $^{12}$CO($J$ = 1--0) maps. The integrated velocity ranges are (a) $V_\mathrm{LSR}$ = $-46$--$-28$ km s$^{-1}$ and (b) $V_\mathrm{LSR}$ = $-15$--0 km s$^{-1}$. The former corresponds the velocity range that is associated with RCW~86, and the latter represents the local molecular cloud velocity. The molecular clouds as shown in Figure \ref{figa1}a is a part of the molecular supershell GS~314.8$-$0.1$-$34 discovered by \cite{2001PASJ...53.1003M}. Figures \ref{figa1}c and \ref{figa1}d show the $Planck$ dust opacity $\tau_{353}$ at the frequency of 353 GHz and the dust temperature map, respectively. We find good spatial correspondence between the CO and $\tau_{353}$ toward the CO $-37$E cloud and filamentary molecular cloud located at the south of the SNR. We also note that the southern shell of RCW~86 shows a high dust temperature $> 20$ K, indicating shock heating.

According to \cite{2017ApJ...838..132O}, total proton column density $N_\mathrm{p}$(H$_2$ + H{\sc i}) can be derived by;
\begin{eqnarray}
N_\mathrm{p}(\mathrm{H}_2 + \mathrm{H}\textsc{i}) = 9.0 \times 10^{24} \; (\tau_\mathrm{353})^{1/1.3} \;\;,
\label{eq8}
\end{eqnarray}
where the non-linear term of 1/1.3 indicates the dust-growth factor discussed by \citet{2013ApJ...763...55R} and \citet{2017ApJ...838..132O}.

Figure \ref{figa2} shows a correlation plot between $W$(CO) and $N_\mathrm{p}$(H$_2$ + H{\sc i}) which was derived by using equation \ref{eq8}. We use the data points of CO with 3 sigma or higher significance, except for the region detected in the local molecular clouds. We perform linear fittings using the {\bf{MPFITEXY}} routine for each dust temperature range, which provides the slope, intercept and the reduced-$\chi^2$ values \citep{2010MNRAS.409.1330W}. We find that using only the data points with dust temperature $< 19.5$ K gives a best-fit with a reduced-$\chi^2 \sim 1.02$ (degree of freedom = 87). We finally obtain that the slope is $\sim 1.0 \times 10^{20}$ cm$^{-2}$ (K km s$^{-1}$)$^{-1}$ and the intercept is $\sim 5.6 \times 10^{21}$ cm$^{-2}$. From equation \ref{eq23}, $N_\mathrm{p}$(H$_2$ + H{\sc i}) can be also written as:
\begin{eqnarray}
N_\mathrm{p}(\mathrm{H}_2 + \mathrm{H}\textsc{i})  &=&   2X \cdot W(\mathrm{CO}) + N_\mathrm{p}(\mathrm{H}\textsc{i}),\\
&\equiv&  (\mathrm{slope}) \cdot W(\mathrm{CO}) + (\mathrm{intercept}).
\label{eq9}
\end{eqnarray}
We therefore obtain the conversion factor $X$ to be $\sim 0.5 \times 10^{20}$ cm$^{-2}$ (K km s$^{-1}$)$^{-1}$.

\end{document}